\documentclass[11pt,twocolumn,amssymb]{revtex4-2}

\usepackage{amsmath}
\usepackage{epsfig}
\usepackage{epic}
\usepackage{color}
\definecolor{bordo}{rgb}{0.8,0.3,0.3}
\definecolor{azul}{rgb}{0.1,0.2,0.7}
\usepackage[english]{babel}
\usepackage{color}
\usepackage{verbatim}
\definecolor{azul}{rgb}{0.1,0.2,0.6} 
\definecolor{verde}{rgb}{0.1,0.5,0.3}
\definecolor{bordo}{rgb}{0.9,0.3,0.3}
\usepackage{lipsum}
\usepackage{slashed}
\usepackage{tikz-cd}
\usetikzlibrary{automata}
\usetikzlibrary{topaths}
\usepackage{pstricks}
\usetikzlibrary{automata,positioning,calc}
\usetikzlibrary{arrows}
\usepackage{circuitikz}
\usepackage{tikz}
\usetikzlibrary{mindmap,trees}
\usepackage{color,xcolor}

\usepackage{mathrsfs}
\usepackage[scr=dutchcal,calscaled=1]{mathalfa}
\usepackage{wasysym}
\usepackage{relsize} 
\usepackage{amsmath,amssymb}
\begin{document}

\title{\Large \sc quantum systems simulatability through classical networks}
\author{M. Caruso}%
\email{mcaruso@fidesol.org}
\affiliation{Fundación I+D del Software Libre, FIDESOL, Granada (18100), España.}
\affiliation{
Campus de Fuentenueva, Universidad de Granada, Granada (18071), Espa\~na.}

\pagestyle{empty}

\begin{abstract}{
We have shown that quantum systems on finite-dimensional Hilbert spaces are equivalent under local transformations. Using these transformations give rise to a gauge group that connects the hamiltonian operators associated with each quantum system. Different quantum systems are connected in such way that studying one of them allows to understand the other. This result can be applied to the field of simulation of quantum systems, in order to  mimic more complicated quantum systems from another simulatable quantum system. Given that there is a bridge that allows to simulate a particular quantum system on this kind of Hilbert spaces using classical circuits we will provide a general scenario to extend this bridge to simulate the time evolution, via Schrödinger equation, of any of these quantum system using classical circuits. This classical systems can be implemented and controlled more easily in the laboratory than the quantum systems.}
\end{abstract}
\vspace{0.3cm}

\date{\today}

\maketitle

\section{Introduction}

In this work, we have developed a procedure to connect a given pair of quantum systems via a local transformation. We describe specifically a map among the respective Hilbert spaces that connect its vector objects (which represent quantum states) and its hamiltonian operators. We will studied the case in which the corresponding Hilbert spaces are finite-dimensional, but this results can be enunciated for infinite, but countable, dimensional Hilbert spaces. This correspondence is a useful tool to map quantum systems in order to study one of them through the other one.

A productive and promising field to apply these ideas may be the \textit{quantum simulations}. At the end of 20th century, R. Feynman asked the following question: \textit{What kind of computer are we going to use to simulate physics?} [...] \textit{I present that as another interesting problem:
to work out the classes of different kinds of quantum mechanical systems which are really intersimulatable $-$which are equivalent$-$} [...] \textit{The same way we should try to find out what kinds of quantum mechanical systems are mutually intersimulatable, and try to find a specific class, or a character of that class which will
simulate everything} \cite{Feynman}.

A quantum simulator is conceived as a controllable system whose aim is to mimic the static or dynamical properties of another quantum system \cite{Georgescu Review}. Today this field can be roughly divided into digital quantum simulations, analog quantum simulations, and a combination of both. The digital quantum simulation, proposed by Lloyd \cite{Lloyd}, deals with the synthesis of a given operator evolution in \textit{quantum gates}. The advantage of this approach is the flexibility, which introduce the quantum error correction and, then, universality. One of its disadvantages is the large number of quantum gates that may be involved in the synthesis, which implies huge technological effort to maintain the coherence of the states. Whereas in the analog quantum simulations there is a \textit{first} quantum system that may be not experimentally easy realizable or controllable and a second quantum system that mimics the first one. The advantage of this approach is the possibility to describe quantum systems in larger Hilbert spaces. One of the main issues concerns to find a quantum system able to mimic certain aspects to simulate the first quantum system. And finally, there is a win-win strategy that consists of a hybrid digital-analog simulations to combine the best of both ideas. \cite{Solano}

Focusing on the digital simulation of hamiltonian dynamics for quantum systems, we see that it is inevitable to deal with a numerable and even finite version of such quantum systems, i.e. the involved Hilbert spaces are finite-dimensional. On the other hand in the analog quantum simulations, we will focus on countable$-$dimensional Hilbert spaces. According to
this, we must study quantum systems on a countable dimensional (denumerable) Hilbert spaces is essentially relevant.

We will show that any two quantum systems on respective Hilbert spaces which are  finite dimensional are connected via a gauge transformation. This includes the case in which any of its corresponding hamiltonian may be time dependent. We intend to deal with the topic of quantum simulation from an alternative perspective,  starting from a more well-known one, in the sense that the last one can be analytically soluble and/or simulatable. We intend to open a way to establish the equivalence class previously mentioned by Feynman \cite{Feynman}.

On the other hand, from another formal equivalence between quantum and classical systems proposed in \cite{caruso} in order to simulate quantum systems through specific circuits, it is possible to use such classical systems which their controllability is simpler than for quantum systems in general, in order to adequately describe its temporal evolution. We will use this equivalence between quantum systems in order to show that a general simulation protocol is possible to implement using classical circuits.


\section{Quantum systems on a denumerable hilbert space}

Reviewing the basics aspects of quantum systems, let us consider a general quantum 
system Q which can be described in a certain $n-$dimensional Hilbert space $\mathscr{H}_n$. The \textit{deterministic} temporal evolution of a quantum system is driven by a hamiltonian operator $H$ (eventually time-dependent) defined on $\mathscr{H}_n$. This operator modifies the vector state $|\psi(t)\rangle$  at time $t{\in\mathcal{T}\subseteq}\mathbb{R}$, by the equation
\begin{equation}\label{schr eqt}
i\partial_t |\psi(t)\rangle=H|\psi(t)\rangle,
\end{equation}
where $\partial_t$ represents the partial time derivative. Note that a partial time derivative is used because $|\psi(t)\rangle$ can be dependent of other quantities. The equation \eqref{schr eqt} is written in natural units, e.g. $\hbar=1$.  A solution of \eqref{schr eqt} is expressed as a $t-$parametrized curve on $\mathscr{H}_n$ $|\psi\rangle{:} \mathcal{T}\longrightarrow \mathscr{H}_n$, this vector curve can be represented using an orthonormal basis of $n-$states: $\pmb{\beta}=\{|\beta_k\rangle\}_{k\in I_n}$, where $I_n{=}\{1,\cdots,n\}$ is the set of the first $n$ natural numbers. The inner product defined in the Hilbert space $\mathscr{H}_n$ allow us to express the state of the system at time $t$, $|\psi(t)\rangle$, in terms of its coordinates in the basis $\pmb{\beta}$ as 
$ \psi_k(t):=\langle \beta_k |\psi(t)\rangle$ where $k\in I_n$.

Note that the \textit{bra-ket} notation  is used to denote the inner product in $\mathscr{H}_n$, $\langle \star | \ast\rangle :\mathscr{H}_n\times \mathscr{H}_n \longrightarrow \mathbb{C}$. Thus, we have a time-parametrized curve on $\mathbb{C}^n$, $
\pmb{\psi}:\mathcal{T}\longrightarrow \mathbb{C}^n$, 
where $\pmb{\psi}$ is written  in terms  of the coordinates of $|\psi\rangle$ in base $\pmb{\beta}$, $\pmb{\psi}{=}(\psi_1,\cdots,\psi_k,\cdots,\psi_n)^\mathfrak{t}$, where $\mathfrak{t}$ is the matrix transposition. Both time-parametrized curves in the abstract Hilbert space and its matrix representation refers to the same quantum system but belongs to different spaces $\mathscr{H}_n$ and $\mathbb{C}^n$. 

Also, the complex vector curve $\pmb{\psi}$ satisfies another version of the equation \eqref{schr eqt}, given by
\begin{equation}\label{schr eqt v2}
id_t \pmb{\psi}(t)=\pmb{H\psi}(t),
\end{equation}
where  $\pmb{H}{\in}\mathbb{C}^{n\times n}$ is a complex matrix that represents the hamiltonian operator $H$ in the basis $\pmb{\beta}$ and whose matrix elements are $H_{kl}=\langle \beta_k |H| \beta_l\rangle$. In this manuscript, we refer to the hamiltonian operator, or hamiltonian matrix simply as hamiltonian.  Note that $d_t$ is a total time derivative is used,  because each coordinate $\psi_k$ depends on time only.

\section{Formal aspects of equivalent quantum systems}\label{sec-4}

We considered a map $\pmb{\Omega}_{\pmb{\omega}}$,  given a nonsingular matrix $\pmb{\omega}$, which transforms a matrix $\pmb{H}\in \mathbb{C}^{n\times n}$ as
\begin{equation}\label{map}
\pmb{\Omega}_{\pmb{\omega}}(\pmb{H}) = \pmb{\omega} \pmb{H} \pmb{\omega}^{-1} +i  (d_t\pmb{\omega}) \pmb{\omega}^{-1},
\end{equation}
where $\pmb{\omega}(t)$ is a $t-$differentiable non$-$singular matrix of $n\times n$, i.e.  $\pmb{\omega}:\mathcal{T}\longrightarrow\mathsf{GL}(n,\mathbb{C})$, also $\pmb{\Omega}_{\pmb{\omega}}\in \mathcal{M}_{n\times n}(\mathbb{C})$.  The map $\pmb{\Omega_\omega}$ \eqref{map} is composed by a similarity transformation of $\pmb{H}$, defined by $\pmb{\omega}$, plus another time-dependent term. The collection of this transformations $\{\pmb{\Omega}_{\pmb{\omega}}: \forall\, \pmb{\omega}\in \mathsf{GL}(n,\mathbb{C})\}$  form a group of local (gauge) transformations, with the composition of maps as a single associative binary operation. The locality of the transformation is due to the $t-$dependence of $\pmb{\omega}$. Note that $\mathcal{M}_{n\times n}(\mathbb{K})$ are the $n\times n$ matrices over the field $\mathbb{K}$.

This kind of mapping was studied in previous works from a pure mathematical point of view for applications to differential equations in complex variables with singular operators \cite{Varadarajan1,Varadarajan2,Varadarajan3}. For a physical point of view the same kind of mapping was presented in \cite{Mustafa1,Mustafa2,Mustafa3} in order to solve particular quantum systems. Respect to that, in this section we studied the possibility to connect any pair of hamiltonian $(H,H')$ operators defined on their respective $n-$dimensional Hilbert spaces $(\mathscr{H}_n,\mathscr{H}_n')$; this hamiltonians are represented by the matrices $(\pmb{H},\pmb{H}')$ eventually time dependent. We proved that there is a non singular matrix $\pmb{\omega}$, $t-$dependent and differentiable, that connect $\pmb{H}$ and $\pmb{H}'$  in this way $\pmb{H}'=\pmb{\Omega}_{\pmb{\omega}} (\pmb{H})$. 

If we composed two transformations $\pmb{\Omega}_{\pmb{\omega}}\circ\,\pmb{\Omega}_{\pmb{\omega}'}$ with $\pmb{\omega}$ and $\pmb{\omega}'$ are nonsingular, we see that  $\pmb{\Omega}_{\pmb{\omega}}\circ\,\pmb{\Omega}_{\pmb{\omega}'}=\pmb{\Omega}_{\pmb{\omega.}\pmb{\omega}'}$, thus \label{commutator}
$[\,\pmb{\omega,\omega}']=\pmb{0}\Longrightarrow [\,\pmb{\Omega_{\omega}},\pmb{\Omega}_{\pmb{\omega}'}]$. So that  $\pmb{\Omega_{I}}=\pmb{I}$, where $\pmb{I}$ is the identity matrix.  If we consider the composed property with $\pmb{\omega}''=\pmb{\omega.\,\omega}'$ such that $\pmb{\omega.\,\omega}'=\pmb{I}=\pmb{\omega}'\pmb{.\,\omega}$ then \label{unique inv} $
\pmb{\Omega_{\omega}}\circ\,\pmb{\Omega}_{\pmb{\omega}'}=\pmb{I}=\pmb{\Omega}_{\pmb{\omega}'}\circ\,\pmb{\Omega}_{\pmb{\omega}}$, we obtain a \textit{unique} inverse of $\pmb{\Omega}_{\pmb{\omega}}$ given by \label{Gammainv} $(\pmb{\Omega_{\omega}})^{{-}1}{=}\;\pmb{\Omega}_{\pmb{\omega}^{-1}}$. 

We demonstrated that for any pair of $n\times n$, eventually $t-$dependent and differentiable, matrices $\pmb{H}$ and $\pmb{H}'$  there exist a non-singular $n\times n$, $t-$dependent and differentiable matrix $\pmb{\omega}$ that connect them. For that we can define the following \textit{equivalence relation}:
\begin{align}\label{equivrel}
\pmb{H}'\sim \pmb{H}\Longleftrightarrow \exists \;\pmb{\omega}: \pmb{H}'=\pmb{\Omega}_{\pmb{\omega}}(\pmb{H}).
\end{align}
From the equivalence relation \eqref{equivrel} then  $\pmb{\omega}$ satisfies the differential equation:
\begin{align}\label{lambdaeq}
id_t\,\pmb{\omega}=\pmb{H}'\pmb{\omega}-\pmb{\omega H}.
\end{align}


First of all, the solution of \eqref{lambdaeq} exists for the trivial cases $\pmb{H}=\pmb{0}$ and $\pmb{H}'=\pmb{0}$, i.e. denoted by $\pmb{\omega}_1$ and $\pmb{\omega}_2$  such that $id_t\,\pmb{\omega}_1{=}\pmb{H}'\pmb{\omega}_1$ \label{l1} and $ id_t\,\pmb{\omega}_2{=}-\pmb{\omega}_2\pmb{H}$\label{l2}. 
Note that we can obtain $(\pmb{\omega}_1,\pmb{\omega}_2)$ as iterative nonsingular solutions following \cite{Magnus}. 
The existence of these solutions $\pmb{\omega}_1$ and $\pmb{\omega}_2$ in terms of the defined equivalent relation \eqref{equivrel} is written as $\pmb{H}' \sim \pmb{0}$ and $\pmb{0} \sim \pmb{H}$, respectively. From  transitivity of the equivalence relation \eqref{equivrel} we have $\pmb{H}'\sim \pmb{H}$. This means that there is a given $\pmb{\omega}$ that $\pmb{H}'=\pmb{\Omega}_{\pmb{\omega}}(\pmb{H})$, and using the composite property of $\pmb{\Omega}_{\pmb{\omega}}$ finally express the solution $\pmb{\omega}$ as a function of the solutions $(\pmb{\omega}_1,\pmb{\omega}_2)$. The equivalences $\pmb{H}' \sim \pmb{0}$ and $\pmb{0} \sim \pmb{H}$  corresponds to $\pmb{H}'=\pmb{\Omega}_{\pmb{\omega}_1}(\pmb{0})$ and $\pmb{0}=\pmb{\Omega}_{\pmb{\omega}_2}(\pmb{H})$ then $
\pmb{H}'=\pmb{\Omega}_{\pmb{\omega}_1}\pmb{(}\pmb{\Omega}_{\pmb{\omega}_2}(\pmb{H})\pmb{)}$ and from the composition rule  applied to $\pmb{H}'=\pmb{\Omega}_{\pmb{\omega}_1\pmb{\omega}_2}(\pmb{H})$, where the \textit{transitive solution} is given by $\pmb{\omega}=\pmb{\omega}_1\pmb{\omega}_2$. The solution $\pmb{\omega}$ built in this way is called a \textit{transitive solution}, or \textit{composite solution}.

We have demonstrated that for any pair of this kind of matrices $(\pmb{H},\pmb{H}')$, there is a nonsingular matrix $\pmb{\omega}$ that connects $\pmb{H}$ and $\pmb{H}'$ through the mapping $\pmb{\Omega_\omega}$, given by the expression \eqref{map}, this is
\begin{equation}\label{Qequiv}
\pmb{H}'=\pmb{\Omega_\omega}(\pmb{H}).
\end{equation}

Suppose now that this pair of matrices  $\pmb{H}$ and $\pmb{H}'$ are the hamiltonian operators of the following differential equations
\begin{align}\label{equations}
id_t\pmb{\psi}=\pmb{H}\,\pmb{\psi},\quad 
id_t\pmb{\psi}'=\pmb{H}'\pmb{\psi}',
\end{align}
finally, from \eqref{Qequiv} and \eqref{equations} we have 
\begin{align}\label{conlc}
\pmb{\psi}'=\pmb{\omega}\pmb{\psi}.
\end{align}
In summary, the connection between  $\pmb{H}$ and $\pmb{H}'$ can be found at the level of the solutions of \eqref{equations}, i.e. $\pmb{\omega}$ connects both hamiltonian matrices via $\pmb{H}'=\pmb{\Omega_\omega}(\pmb{H})$ \eqref{Qequiv} and also both solutions via $\pmb{\psi}'=\pmb{\omega}\pmb{\psi}$ \eqref{conlc}. Of course, if the natural units are not used, will be able to define a similar map $\pmb{\Omega_\omega}$, multiplying  the second term on the right side of expression \eqref{map}  by $\hslash$.

An important comment about the map $\pmb{\Omega_\omega}$: if $\pmb{\omega}$ is a unitary matrix (i.e. $\pmb{\omega}^{\dagger}=\pmb{\omega}^{-1}$, where $\dagger$  corresponds to a complex transposition operation), then $\pmb{\Omega_\omega}$ is an endomorphism over the space of the self-adjoint operators (each of those are defined over the respective Hilbert spaces that have the same dimension $n$). This situation corresponds to a mapping of self adjoint hamiltonian operators associated to closed quantum systems.



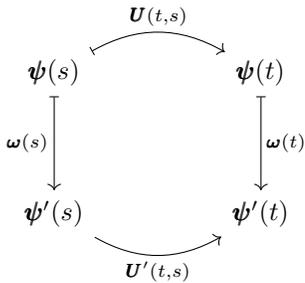
\begin{figure}[h!]
\[
\begin{tikzcd}
\pmb{\psi}(s) \arrow[rr, "{\pmb{U}(t,s)}", maps to, bend left] \arrow[dd, "\pmb{\omega}(s)"', maps to] &  & \pmb{\psi}(t) \arrow[dd, "\pmb{\omega}(t)", maps to] \\
&  &                                                         \\
\pmb{\psi}'(s) \arrow[rr, "{\pmb{U}'(t,s)}"', bend right]                                              &  & \pmb{\psi}'(t)                                      
\end{tikzcd}\]
\caption{\footnotesize{This commutativity diagram shows how is the composition of transformations between equivalent quantum systems. The commutativity comes from the existence of the inverse of  $\pmb{\omega}(x)$ or the inverse of $\pmb{U}(y,z)$ for all $x,y,z$.}}\label{diagrama}
\end{figure}

In the general case these hamiltonians $(\pmb{H},\pmb{H}')$ are not necessarily hermitian and even they can be time-dependent. We can express the evolution operator of each quantum systems $(\pmb{U},\pmb{U}')$ related via $\pmb{\omega}(t)$ as:  $\pmb{U}'(t,s)=\pmb{\omega}(t)\,\pmb{U}(t,s)\,\pmb{\omega}^{-1}(s)$, for all $t,s\in \mathbb{R}$. In  Figure \ref{diagrama} a  commutative diagram shows how is the composition of this transformation.  As we said for the case of $(\pmb{H},\pmb{H'})$ are hermitians we have 
\begin{equation}\label{hibrid}
\pmb{U}'(t,s)= \pmb{\omega}(t)\,\pmb{U}(t,s)\,\pmb{\omega}^{\dagger}(s).
\end{equation}
where $\pmb{\omega}$ is an unitary matrix. 

A general strategy to simulate the \textit{arrival} system Q$'$ consists the selection on the \textit{departure} system  Q and the connection between them, given by $\pmb{\omega}$. The departure system Q is chosen in order to simulate it analogically or digitally. 
The bridge between Q and Q$'$, namely  $\pmb{\omega}$,  is not necessarily associated with any particular quantum system. For this reason the operator $\pmb{\omega}$ will be digitally simulated from a synthesis in quantum gates \cite{Giles Selinger}, which is an universal method. Therefore the expression \eqref{hibrid} could be implemented through a hybrid kind of simulation, that involves a digital-analog protocol for its implementation, as an indispensable requirement \cite{Solano}.

\section{A connection with electrical networks}

In previous works Rosner \cite{Rosner} has proposed an interesting analogy between particular quantum system and a classical system of electrical oscillators. This specifical proposal has been formalized and experimentally performed in \cite{caruso0,caruso}. In this paper we will do two things: one is to generalize this formalization in order to include quantum systems over  an arbitrary dimensional Hilbert space. On the other hand, making use of the fact that all these quantum systems are equivalent via $\pmb{\Omega_\omega}$, the route by which it is possible to implement the simulation of quantum systems via classical circuits is well defined. 

Given that the equivalence between quantum systems, we can consider without loss of generality, a time-independent and self-adjoint hamiltonian operator, represented by a constant hermitian matrix $\pmb{H}$. Let's start by rewriting \eqref{schr eqt v2} using the \textit{decomplexification}  procedure \textit{a la} Arnold \cite{Arnold,Arnold2}, also called \textit{realification} of the space $\mathbb{C}^n$ through the operator $\mathfrak{D}:\mathbb{C}^{n}\longrightarrow\mathbb{R}^{2n}$, explicitly defined as $\mathfrak{D}(\pmb{\psi}):=\pmb{(}\mathrm{Re}(\pmb{\psi}),\mathrm{Im}(\pmb{\psi})\pmb{)}^\mathfrak{t}=\pmb{(}\pmb{\varphi}_1,\pmb{\varphi}_2\pmb{)}^\mathfrak{t}$; which separates the real and the imaginary part of the complex $\pmb{\psi}$ and $\mathfrak{t}$ denotes the matrix transposition.

The equation \eqref{schr eqt v2} take the form of two separate equation for real and imaginary part of $\pmb{\psi}$.
\begin{align}
	\begin{aligned}
	\label{realimag}
		\pmb{\dot{\varphi}}_1 &= \pmb{H}_2\, \pmb{\varphi}_1 - \pmb{H}_1\, \pmb{\varphi}_2  \\
		\pmb{\dot{\varphi}}_2 &= \pmb{H}_1\, \pmb{\varphi}_1 + \pmb{H}_2\, \pmb{\varphi}_2, 
	\end{aligned}
\end{align}
where  $\pmb{H}_1$ and  $\pmb{H}_2$ are the real and imaginary part of the matrix $ \pmb{H}$, respectively. After a standard decoupling procedure one gets the equations
\begin{equation} \label{soe}
\ddot{\pmb{\varphi}}_l(t) +\pmb{A}_\mathfrak{q} \,\dot{\pmb{\varphi}}_l(t) +
\pmb{B}_\mathfrak{q} \,\pmb{\varphi}_l(t)=\pmb{0},
\end{equation}
valid for $l=1,2$, it is clear that even if both the real and imaginary part of $\pmb{\psi}$ verify the same equation, one cannot leave out one of them because the solution of \eqref{soe} implies the knowledge of the initial conditions. The $dots$ notations refers to the time derivative and the real matrices $\pmb{A}_\mathfrak{q}$ and $\pmb{B}_\mathfrak{q}$ depends on $(\pmb{H}_1,\pmb{H}_2)$  \cite{caruso}, the subscript refers to the \textit{quantum} origin of this matrices.

Let us now go to a classical system. We start with a
system of linear differential equations of second order, entirely similar to  \eqref{soe}
\begin{equation}\label{A B}
\ddot{\pmb{q}}(t)+\mathbf{A}\:\dot{\pmb{q}}(t)+\mathbf{B}\:\pmb{q}(t)=\pmb{0}
\end{equation}
with $\pmb{q}{\in}\mathbb{R}^n$ are the generalized coordinates, and $\mathbf{A}, \mathbf{B}{\in}\mathcal{M}_{n\times n}(\mathbb{R})$. Following  \cite{caruso0,caruso} this kind of equations can be performed through a classical electric network. We are particularly interested in lumped element model circuits where voltage and current depend only upon time.

The corresponding dynamics of an electric network is defined by the appropriate use of the Kirchhoff rules that take care of the topology of the network. We restrict our analysis to \textit{passive} networks, where the energy provided by an external source is non negative. The network has exactly $n-$ports: pairs of terminals that allow to exchange energy with the surrounding and have a given voltage and current. One has the possibility of choosing the voltage or the current as the representative state variable of the excitation or the response of the network. We call $\pmb{v}$ the vector corresponding to the port voltage and $\pmb{i}$ the port current, in \cite{caruso0, caruso} we choose the port voltage $\pmb{v}$ to implement a particular electric circuit.

\begin{figure}[h!]
	\centering
	\resizebox{0.4\textwidth}{!}{%
		\begin{tikzpicture}
			\path[mindmap,concept color=black!70,scale=1.3]
			node[concept,minimum size=4.5cm, color=black!70 ,text=white,scale=1.3] 
			{\Huge $\pmb{\mathcal{N}}$}
			[clockwise from=90]
			child[concept color=gray!50!, text=black] 
			{node[concept] {\Huge $\pmb{\mathcal{N}_1}$}}  
			child[concept color=gray!50!,text=black,grow=35] 
			{node[concept] {\Huge $\pmb{\mathcal{N}_2}$}}
			child[concept color=gray!50,text=black,grow=-90] { node[concept]
				{\Huge $\pmb{\mathcal{N}_k}$}}
			child[concept color=gray!50!, text=black,grow=145] 
			{node[concept] {\Huge $\pmb{\mathcal{N}_n}$} };
			\draw[gray,ultra thick,circle,fill=gray] (5.4cm,0.36cm) circle(0.15cm);
			\draw[gray,ultra thick,fill=gray] (4.87cm,-2.53cm) circle(0.15cm);
			\draw[gray,ultra thick,fill=gray] (2.85cm,-4.87cm)   circle(0.15cm);
			\draw[gray,ultra thick,circle,fill=gray](-5.4cm,0.36cm) circle(0.15cm);
			\draw[gray,ultra thick,fill=gray] (-4.87cm,-2.53cm) circle(0.15cm);
			\draw[gray,ultra thick,fill=gray] (-2.85cm,-4.87cm)   circle(0.15cm);
		\end{tikzpicture}
	}
	\caption{\footnotesize{There are $n$ dipole networks, denoted by $\{\mathcal{N}_1,{\cdots},\mathcal{N}_k,{\cdots},\mathcal{N}_n\}$ interconnected through an \textit{interaction} network $\mathcal{N}$.}}\label{nTopology}
\end{figure}
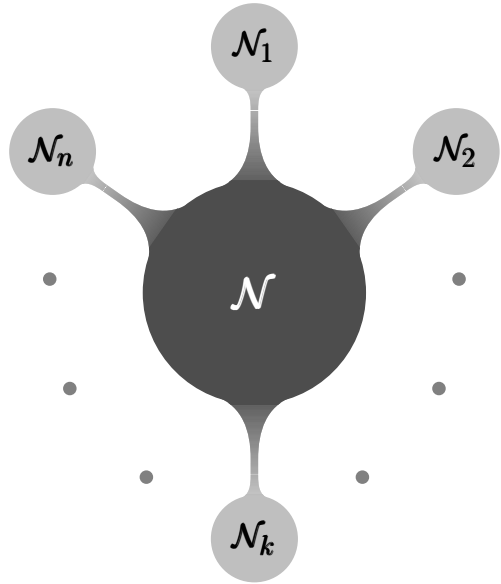

The analysis provided in \cite{caruso0} of the time evolution of electric circuits results in a system of linear differential equations as \eqref{A B}.   Using the \textit{synthesis} methods from \cite{Bala,Carlin} of all electric networks in a given family also developed in \cite{caruso0} the  proposed topology of this kind of electric network is represented in the figure \ref{nTopology}. There are $n-$dipole networks of kind $L\Vert C$ tandem circuit interconnected through an network $\mathcal{N}$.

We consider without loss of generality that the $n-$port electric network $\mathcal{N}$ admit an \textit{admittance} representation with admittance matrix $\pmb{Y}$, an $L\Vert C$ tandem circuit is connected to each port, the  elements of $\pmb{Y}$ are obtained from the general method of $n-$port network synthesis method presented in \cite{Carlin}. Using the Laplace transformation, moving from the temporal domain to the $s-$domain variable it is possible to obtain the elements of $\pmb{Y}$ in order to obtain a differential equation similar to \eqref{A B}, i.e. $\pmb{Y}(s)=\pmb{\alpha}+\frac{1}{s}\pmb{\beta}	$ contains resistors and inductors only, in the constant matrices $\pmb{\alpha}$ and  $\pmb{\beta}$, respectively.
 
A concrete construction of such circuits with their corresponding experimental measurements can be found in \cite{caruso}. But in general from the configuration given by the figure \ref{nTopology} each port voltage $v_k$ works as the real, or imaginary, part of the $k-$component of the wave function $\pmb{\psi}$. From this generalization the classical matrices in \eqref{A B} are given by 
$\pmb{A}=\pmb{C}^{-1}\pmb{\alpha}$ and $\pmb{B}=\pmb{\omega}^2_0+\pmb{C}^{-1}\pmb{\beta}$, where $\pmb{C}=diag(C_1,\cdots,C_n)$ and $\pmb{L}=diag(L_1,\cdots,L_n)$ are the diagonal matrices which contain the capacitors  and inductors of each of the $n-$dipole networks $\{\mathcal{N}_k\}_{k=1,\cdots n}$. Thus the matrix $\pmb{\omega}^2_0:=(\pmb{LC})^{-1}$ contains the proper frequencies of each $L\Vert C$ port tandem connected, see figure \ref{nTopology}.

\section{Conclusion and final observations} \label{sec-5}

The aim of present work it was to show that there is a way to modify the behavior of a known quantum system, in order to get information of another quantum system that, at least, has a difficulty to be resolved directly.


In summary, we have shown how for a given pair of quantum systems, finite-dimensional Hilbert spaces and its respective hamiltonian: $(\mathscr{H},\pmb{H})$ and $(\mathscr{H}',\pmb{H}')$ they could be linked via gauge (local) transformations $\pmb{\omega}$, that allow us to obtain $\pmb{H}'$ from $\pmb{H}$, via $\pmb{\Omega_\omega}$. 



Respect to the simulation of a quantum system Q$'$ we search for some other system that imitates the behavior of Q as well as possible. In others words, we must perform a \textit{casting call} of quantum systems or \textit{actors} which can be very limited, because it is a hard task to find another Q one to simulate Q$'$. We wanted to use this equivalence between quantum systems to simulate \textit{another} quantum system connected with  Q$'$. But when we said \textit{another}, we want to say \textit{any other} quantum system which is connected with Q$'$ through $\pmb{\omega}$. The map $\pmb{\Omega_\omega}$ applied to a given hamiltonian $\pmb{H}$  in \eqref{map}  works as \textit{makeup} that allows any \textit{actor} Q, to  simulate the first quantum system Q$'$, a priori, if Q is connected with Q$'$ through $\pmb{\omega}$.  Following the metaphor, the equivalence between this quantum systems expands that catalogue of actors that can make a good performance in order to mimic another quantum system and becoming that casting call, a priori more efficient. 


A final comment in this regard could be the implementation of the formal equivalence between quantum and classical systems proposed in \cite{caruso} in order to simulate quantum systems through specific circuits. The advantage of using such classical systems is that their controllability is simpler than for quantum systems in general, in order to adequately guide its temporal evolution. This implementation open the possibility to expands this catalogue  of actors capable of simulating the quantum system $Q$  even more with classical actors, who usually do not play that role.

We are convinced that this work enriches the field of quantum simulations and paves the way for new protocols in hybrid (digital-analog) simulation. But also opens the way to new approach to solve
quantum systems. Studying the solution of a simple one in order to obtain the solutions for another
quantum system, eventually a more complicated one. This approach allows simulating quantum systems establish a strong connection between them and so build a dictionary for interpreting concepts of a theory on the other.

\section*{Acknowledgments}

We thank to FIDESOL for the support and recall also the anonymous readers for their constructive criticism to this work.

\section*{Competing interest}
The author declares that there are no competing interests.

\vspace*{0.5cm}

\end{document}